\documentclass{mn2e}
\usepackage{epsfig}

\def\le{{L_{\rm Edd}}}
\def\msun{{\rm M_{\odot}}}

\def\rp{{R_{\rm ph}}}
\def\rs{{R_{\rm s}}}
\def\mo{{\dot M_{\rm out}}}
\def\me{{\dot M_{\rm Edd}}}
\title[Black Hole Winds]
{Black Hole Winds}
\author[A.R.~King and K.A. Pounds]{
A.R.~King$^{1, 2}$ and K.A. Pounds$^1$\\
1. Department of Physics and Astronomy, University of Leicester,
Leicester, LE1~7RH, \\
2. Harvard--Smithsonian Center for Astrophysics, 60 Garden St,
Cambridge, MA 02138, USA
}

\begin{document}

\maketitle

\begin{abstract}
We show that black holes accreting at or above the Eddington rate
probably produce winds which are optically thick in
the continuum, whether in quasars or X--ray
binaries. The photospheric radius and
outflow speed are proportional to $\mo^2$ and $\mo^{-1}$
respectively, where $\mo$ is the mass outflow rate. The outflow momentum rate
is always of order $L_{\rm Edd}/c$. Blackbody emission
from these winds may provide the big blue bump in some quasars and AGN,  
as well as ultrasoft X--ray components in ULXs.
\end{abstract}

\begin{keywords}
black hole physics -- accretion --  X--rays: galaxies -- X--rays: binaries --
quasars: general

\end{keywords}

\section{Introduction}

Recent {\it XMM--Newton} observations of bright quasars 
(Pounds et al., 2003) have revealed strong evidence for 
intense outflows at mass rates comparable to Eddington. 
Such outflows 
closely resemble those recently inferred in a set of 
ultraluminous X--ray sources (ULXs) with extremely soft spectral 
components (Mukai et al, 2003; Fabbiano et al., 2003). 
In the quasar PG1211+143 Pounds et al. (2003) 
find blueshifted X--ray absorption lines which
show that the outflow has high velocity ($v \sim 0.1c$)
The X--ray absorption columns in the quasar outflows are very large
($\sim 10^{24}$~cm$^{-2}$) suggesting that they may be Compton thick
at small radii. Pounds et al. (2003) were able to use this information
and the ionization state of the absorbing material to infer a mass
outflow rate $\mo \sim 1.6\msun~{\rm yr}^{-1}$. This is close to the
Eddington accretion rate $\me$ for this object.

Pounds et al. further showed that mass conservation
strongly suggests that any outflow with $\mo \sim \me$ is likely to 
be optically thick to electron scattering, with a
photospheric radius $R_{\rm ph}$ of order a few 10s of Schwarzschild
radii $\rs = 2GM/c^2$, where $M$ is the black hole mass. 
The generic nature of the outflow
characteristics at supercritical accretion rates (see e.g. eq. \ref{phot})
below) strongly suggests
that these outflows may be a widespread phenomenon, not only in currently
observed systems such as quasars and ULXs, but also in the growth of
supermassive black holes in the centres of galaxies in the past.

For the QSO PG1211+143 Pounds et al (2003) found 
a photospheric radius $\rp \sim 150\rs$
from their estimate of $\mo/\me$. Although unremarked in that paper,
the outflow velocity independently found from the
X--ray absorption lines is very close to the escape velocity from this 
radius. Further, the outflow momentum rate $\mo v$ is of precisely the same
order as the radiation momentum rate $\le/c$. 

We show here that these features are just as expected in an optically
thick wind driven by continuum radiation pressure. We use this to give 
simple scalings for the outflow velocity and photospheric radius $v,
\rp$ in terms of $\mo/\me$.

\section{Outflows from Eddington--Limited Accretors}

We outline here a simple theory of outflows from black holes accreting
at rates comparable with the Eddington value
\begin{equation}
\dot M_{\rm Edd} = {4\pi GM\over \eta \kappa c}.
\label{medd}
\end{equation}
Here $\eta c^2$ is the accretion yield from unit mass, and
$\kappa$ is the electron scattering opacity. 
We assume that the outflow is radial, in a double cone 
occupying solid angle $4\pi b$,
and has constant speed $v$ for sufficiently large radial distance
$r$. We will justify the second assumption later in this Section.
Mass conservation implies an outflow density
\begin{equation}
\rho = {\mo\over 4\pi vbr^2}.
\label{cons}
\end{equation}
The nature of the outflow depends on $b$. If $b \sim 1$ we can neglect
scattering of photons from the sides of the outflow, while for $b <<
1$ this process is dominant. For completeness we first briefly revisit
the case $b \sim 1$ (cf Pounds et al. 2003) 

The electron scattering
optical depth through the outflow, viewed from infinity down to radius
$R$, is
\begin{equation}
\tau = \int_R^\infty\kappa\rho{\rm d}r = {\kappa\mo\over 4\pi vbR}.
\label{tau}
\end{equation}
From (\ref{medd}, \ref{tau}) we get
\begin{equation}
\tau = {1\over 2\eta b}{R_{\rm s}\over R}{c\over v}{\mo\over
\dot M_{\rm Edd}}.
\end{equation}
Defining the photospheric radius $R_{\rm ph}$ as the point $\tau = 1$ gives
\begin{equation}
{R_{\rm ph}\over R_{\rm s}} = {1\over 2\eta b}{c\over v}{\mo\over \dot
M_{\rm Edd}}\simeq {5\over b}{c\over v}{\mo\over \dot M_{\rm Edd}}
\label{phot}
\end{equation}
where we have taken $\eta \simeq 0.1$ at the last step. Since $b \leq
1, v/c < 1$ we see that $R_{\rm ph} > R_{\rm s}$ for any outflow rate
$\mo$ of order $\dot M_{\rm Edd}$, that is, such outflows are
Compton thick.

If instead $b << 1$, photons typically escape from the side of
the outflow rather than making their way radially outwards through all
of it. Almost all of the photons escape in this way within 
radial distance $r = R_{\perp}$ 
where the optical depth across the flow
\begin{equation}
\tau_{\perp} \simeq \kappa\rho(r) b^{1/2}r
\label{tauperp}
\end{equation}
is of order unity. Thus
\begin{equation}
{R_{\perp}\over R_{\rm s}} = {1\over 2\eta b^{1/2}}{c\over v}{\mo\over \dot
M_{\rm Edd}}\simeq {5\over b^{1/2}}{c\over v}{\mo\over \dot M_{\rm Edd}},
\label{rperp}
\end{equation}
and we again conclude that the outflow is Compton thick for $\mo
\sim \me$.

This conclusion evidently implies that much of the emission from such
objects will be thermalized and observed as a softer spectral
component (see eq \ref{teff} below). The observed harder X--rays must
presumably either be produced near the `skin' of the outflow (i.e. at
moderate $\tau$), or result from shocks within the outflow. In both
cases their total luminosity must be lower than that of the
thermalized soft component.

We now investigate how the outflow is driven.
Since the wind is Compton thick most of the photons have scattered
and thus on average given up their original momentum to the outflow.
Outside the radius $R_{\rm ph}$ or $R_{\perp}$ the photons decouple
from the matter and there is no more acceleration. This justifies our
assumption that $v$ is constant for large $r$, and it is
self--consistent to use the assumption to integrate inwards to 
$R_{\rm ph}$ or $R_{\perp}$. 

To ensure that the matter
reaches the escape speed we require the radii $R_{\rm
ph}, R_{\perp}$ to lie close to the escape radius $R_{\rm esc}$, i.e.
\begin{equation}
R_{\rm ph}, R_{\rm \perp} \simeq R_{\rm esc} = {c^2\over v^2}R_{\rm s}.
\label{esc}
\end{equation}
From this equation and (\ref{phot}, \ref{rperp}) we find
\begin{equation}
{v\over c} \simeq {2\eta f\me\over \mo},\ \  R_{\rm ph,\ \perp}
\simeq \biggr({\mo\over 2\eta f\me}\biggr)^2
\label{scaling}
\end{equation}
where $f = b, b^{1/2}$ in the two cases $b \la 1, b << 1$. 
We can write these formulae more compactly as
\begin{equation}
{v\over c} = {2f\le\over \mo c^2}
\label{v}
\end{equation}
\begin{equation}
{R_{\rm ph,\ \perp}\over \rs} = \biggl[{\mo c^2\over 2f\le}\biggr]^2,
\label{r}
\end{equation}
We note
that $f \sim 1$ except for very narrowly collimated outflows ($b
\la 10^{-2}$).

An immediate consequence of (\ref{v}) is 
\begin{equation}
\mo v = 2f{\le\over c},
\label{mom}
\end{equation}
i.e. the momentum flux in the wind is always of the same order as that
in the Eddington--limited radiation field, as expected for an
Compton thick wind driven by radiation pressure. The energy flux
(mechanical luminosity) of the wind is lower than that of the
radiation field by a factor of order $v/c$:
\begin{equation}
\mo{v^2\over 2} = {v\over c}f\le = {2(f\le)^2\over \mo c^2}.
\label{en}
\end{equation}

\section{The Blackbody Component}

Since the outflow is Compton thick for $\mo \sim \me$, 
much of the accretion luminosity generated 
deep in the potential well near $\rs$ must emerge as blackbody--like 
emission from it. If $b \sim 1$ the quasi--spherical radiating area is
\begin{equation}
A_{\rm phot} = 4\pi b\rp^2 
\end{equation}
If instead $b << 1$ the accretion luminosity emerges from the curved
sides of the outflow cones, with radiating area 
\begin{equation}
A_{\perp} = 2\pi\rp^2b^{1/2}
\end{equation}
We can combine these two cases as
\begin{equation}
A_{\rm ph, \perp} = 4\pi g\biggl[{\mo c^2\over 2\le}\biggr]^4\rs^2
\label{area}
\end{equation}
with $g(b) = 1/b, 1/2b^{1/2}$ in the two cases.
Again $g \sim 1$ unless $b \la 10^{-2}$, so the areas are similar in
the two cases. However we note that the radiation patterns differ. In
particular if $b$ is small radiation is emitted over a
wider solid angle than the matter. 
Numerically we have
\begin{equation}
A_{\rm ph, \perp} = 2\times 10^{29}g\dot M_1^4M_8^{-2}~{\rm cm}^2,
\label{aeff}
\end{equation}
where $\dot M_1 = \mo/(1\msun~{\rm yr}^{-1}), M_8 = M/10^8\msun$.
The effective blackbody temperature is
\begin{equation}
T_{\rm eff} = 1\times 10^5g^{-1/4}\dot M_1^{-1}M_8^{3/4}~{\rm K}.
\label{teff}
\end{equation}
Clearly such a component is a promising candidate for the soft excess
observed in many AGN and ULXs.

\section{Discussion}

We have investigated supercritical accretion ($\dot M \ga \me$)
on to black holes. We assume that the excess matter is ejected, and
have shown that the resulting outflow is Compton thick. The only
alternative to this is to assume that the hole is able to accrete most
of the mass at low radiative efficiency. However recent numerical 
simulations suggest (e.g. Stone \& Pringle, 2001) that in this case
most of the mass is ejected by the black hole rather than accreted.

We therefore expect that any black hole
accreting at a rate $\dot M \ga \me$ will
show a strong Compton thick outflow, with effective
photosphere of size a few 10s of $\rs$, scaling as $\mo^2$
The outflow velocity is of order the escape
velocity from this photosphere, and scales as $\mo^{-1}$. Observations
of the QSO PG1211+143 strongly support this picture.

In this picture 
much of the accretion energy must be emitted from the photosphere
with typical temperature given by (\ref{teff}). Mukai et
al. (2003) and Fabbiano et al. (2003) use this to explain the very
soft X--ray spectral components found in some ULXs, although they were
forced to assume an outflow velocity rather than estimating it
self--consistently as here. This is in line with the
suggestion (King et al., 2001; King, 2002) that 
ULXs are X--ray binaries where the current accretion rate 
$\dot M \ga \me$. The particular
ULX discussed by Fabbiano et al. (2003) has a blackbody 
temperature of order
$10^6$~K, and eqn (\ref{teff}) shows that a $10\msun$ black hole would need
an outflow (and thus mass transfer) rate of order $10\me \sim
10^{-6}\msun~{\rm yr}^{-1}$. 
This is
not extreme for thermal timescale mass transfer in a
massive X--ray binary (`SS433--like') 
or for a bright transient (`GRS1915+105--like'), the two situations
envisaged for ULX behaviour by King et al. (2001) and King (2002). 
A similar
value appears to hold for the ULX discussed by Mukai, where it is
noticeable that the blackbody luminosity remains constant while the
temperature changes by factors $\sim 2$. 

An important question raised by our work is whether the outflow velocities
discussed here can be identified with those of 
the jets seen in both classes of
ULX. If so, the observed jet velocities would give direct information
about $\mo/\me$ and thus the accretion rate through eq (\ref{v}). Thus
the $v = 0.27c$ jets seen in SS433 would imply a mass transfer rate
$\sim 5\me$. This idea needs caution, as the jet may simply
represent the fastest part of the outflow, rather than carrying most
of the outflowing mass. In particular the jets are known to be
inhomogeneous blobs rather than continuous outflow.
Moreover jets are seen in systems where the
luminosity is below or not significantly higher than $\le$.
Interestingly, it is clear that the radiation pattern
in the microquasar GRS1915+104 (isotropic luminosity
$\simeq 4 \le$, cf King, 2002 
is indeed wider than the matter outflow, i.e.
$b({\rm radiation}) > b({\rm outflow})$. This would agree with the
suggested radiation pattern for the case $b({\rm outflow}) << 1$
discussed above.

The very general nature of the arguments
presented here suggests that outflows may be the seat of ultrasoft
components in ULXs, and of the big blue bump in AGN and QSOs accreting at 
close to the Eddington limit. It will be important to study how they
interact with their surroundings.

\section{Acknowledgments} 

This work was completed at the Center for Astrophysics,
and ARK thanks members of its staff, particularly Pepi Fabbiano and
Martin Elvis, for stimulating discussions and warm hospitality.
ARK gratefully acknowledges a Royal Society Wolfson Research
Merit Award.

\end{document}